# Architecture of Network Camera Photo Authentication Scheme using Steganography Approach

Ahmad M. Nagm [1]
[1]Electrical Engineering, Faculty of Engineering, Al Azhar University, Egypt
Khaled Y. Youssef [2]
[2] Electronic and Communications, Suez University, Egypt
Mohammad I. Youssef [3]
[3]Electrical Engineering, Faculty of Engineering, Al Azhar University, Egypt

**Abstract**— The aim of integrity protection process is not only to secure the send message, but also ensures that the original message is not modified during sending the message. In this paper, a novel proposed model is proposed to verify that the sent message was not modified. The proposed model starting with encrypting the original message and applying one of the hash functions on the message. The ciphered message, private key, and the output of hash function was reshaped to apply Discrete Cosine Transform (DCT), Zig-zag ordering, and Huffman coding to obtain a compressed text, this compressed text is then hidden in an image. The same steps with a reverse order are applied on the receiver to obtain the decrypted message and then apply the same hash function to verify that the message was not changed. Different experiments were conducted and the results demonstrated that the proposed model achieved high integrity protection.

——————————— ◆ ———————————

## 1 INTRODUCTION

The history of the photographic image as a source of information has been questioned since the invention of photography. The credibility of news and documentary photography depend on the faithful recording of the events. The integrity issue of such type of photography is increasing dramatically with the introduction of digital technology. Many people are now living in a world where billions of images are uploaded on social media that in turn represents further challenges with regard to the credibility of images [1]. Moreover, the digital revolution has transformed photography in more ways than we may have realized before. For ex-ample, in traditional former analogue photography, the camera is understood as a picture-making device. It was believed that the picture is formed in the instant you click the shutter on the camera. You might modify the color of the photo after the fact but the main characteristics of the photo were defined at the initial snapshot of the camera shutter. However, in the digital era, the camera is no more perceived as a picture making device but rather a data-collection device that is capturing as much data as possible about the scene. The data collection part should be followed by an advanced computational technique to process the data into the final image. Hence, the term original copy needs a redefinition because what is defined at the time of capture is not a fully formed picture [1], [2]. In addition, the introduction of network cameras makes the situation more difficult in terms of integrity and credibility as the connection of cameras to the internet make the photo more vulnerable to attacks with a potential impact on integrity [2]. In cameras, JPEG processes are done in-camera following the manufacturer's algorithms while RAW conversion is outside of the camera under the control of the photographer or editor. Accordingly, Digital manipulation inside the camera should put a signature based on how images are captured in camera and then post-processed outside the camera [2], [3]. This paper discusses the outcomes of a research held to study the methodology to make image authentication plus hiding the digital signature in the image using steganography approaches together with appropriate algorithms for selection, classification, and hiding of the image authentication patterns. The resultant image out of the highlighted processing stage could be labeled the original message. The paper shall answer the question of what is the original message versus the challenges of the digital revolution and high networking capabilities [3-5].

# 2 MECHANISM-METHODOLOGY oF SECURITY oF COMPRESSED DIGITAL IMAGE

## 2.1 Data Integrity Algorithms
Integrity protection has many types of algorithms such as like cryptography hash function, e.g. Secure Hash Algorithms (SHA) family and MD algorithms. Also, there is also message authentication code and digital signature. In our model, one of these algorithms is used the cryptography algorithm designed to authenticate the identity of entities.

## 2.2 Cryptography Algorithms
The aim of the cryptography technique is to secure information from malicious intruder by conceal contents of data, protect blocks of data, or authenticate identity for entities from active attacks which modify the contents of data, masquerade, replay, and denial of services. The cryptography algorithm has the following elements:
- Plaintext: This is the original message or data as input to the algorithm.
- Encryption algorithm: The encryption algorithm performs various processes on the plaintext.
- Cipher-text: This is the scrambled message produced as output.
- Decryption algorithm: This is essentially the encryption algorithm run in reverse order. This algorithm takes the ciphertext and the secret key and produces the original plaintext.

Encryption data algorithms either symmetric (ciphered the contents of blocks or streams of data of any size as a messages or files do that by using DES-3DES-AES algorithms when the sender and receiver using the same key) or by asymmetric algorithms (for ciphered small blocks of data like an encryption keys or hash function values and digital signature by RSA-Elliptic curve algorithms if the sender and receiver use different keys) [6].

## 2.3 Steganography Techniques
The word steganography was derived from two Greek words "stegos" which meaning "cover" and "grafia" that meaning "writing" defining it as "covered writing". Transmitting information in a secured form as innocent as an encrypted credit card number to an online-store as insidious as a terrorist plot to hijackers [6]. Steganographic messages are often first encrypted by some traditional means, after that a cover image is modified in some way to contain the encrypted message. The goal of steganography technique is to hide the messages inside other harmless messages, which means 'writing in hiding'. In other words, is to hide data in a cover media such as an image, audio file, video stream, so that others will not be able to notice it [7]. While cryptography is about protecting the content of messages, steganography is about concealing their very existence. Due to steganography can be achieved using any cover media we used hiding data in digital images.

A few key properties that must be considered when creating a digital data hiding system are:
Imperceptibility: is the property in which a person should be unable to distinguish the original and the stego-image.
Embedding Capacity: Refers to the amount of secret information that can be embedded without degradation of the quality of the image.
Robustness: Refers to the degree of difficulty required to destroy embedded information without destroying the cover image.

## 2.4 Digital Image Compression Techniques
The main idea of data compressing is reducing the data correlation and replacing them with simpler data form. Then we will discuss the method that is commonly used in image/video compression. We will introduce quantization and entropy coding. After reducing data correlation, the amounts of data are not really reduced. We use quantization and entropy coding to compress the data. Image compression techniques have two main types, namely, lossy and lossless compression techniques. A lossy compression reduces a file by eliminating redundant information. In this method, the user makes a tradeoff between file size and image quality and the some of the data are lost during the compression process. On the other hand, in a lossless compression, there is no data of the original file are lost. In other words, all the information can be completely restored [10].

## 2.5 Overview on Security of Compressed Digital Image
Cryptography is providing high security for digital image and protects it from hacking or spying. There are various techniques to protect the confidentiality, integrity, and authenticity of images to make images more secure. Image encryption is one of the most important applications in transferring images through the internet and cellular phones [8]. In this application, the symmetric encryption which encrypt and decrypt the message are performed using the same key. This process also including (1) substitution techniques such as Caesar Cipher, Monoalphabetic Ciphers, Hill Cipher, Polyalphabetic Ciphers, and One-Time Pad, (2) Transportation techniques (rail fence cipher)., and (3) Rotor machines.
A classic stream cipher encrypts a digital data stream one bit or one byte at a time (autokeyed VigenÃlreciphe- Vernam-chiper). A block cipher is an encryption/decryption method in which a block of plaintext, i.e. original message, is treated as a whole and used to produce a ciphertext block of equal length. The most widely used symmetric cipher is the Data Encryption Standard (DES) and it is replaced by Advanced Encryption Standard (AES) [9]. Asymmetric encryption which encryption and decryption are performed using the different keys, one for a public key and another for a private key. It is also known as public-key encryption. The most widely used public-key cryptosystem is RSA. The difficulty of attacking RSA is based on the difficulty of finding the prime factors of a composite number.

Steganography technique is conceal contents of data (secret message) by inserting it into another data (cover-medium) to be stego-medium with key about certain methods and cryptography technologies are combined to satisfy the need for privacy on the Internet. The applications of information hiding systems are widely used in different applications such as military, intelligence agencies, online elections, internet banking, medical-imaging and so on. Steganography equation is defined as follows: Stego-medium = Cover-medium + Secret-message + Stego-key. There exist two types of materials in steganography, namely, message and carrier, the message is the secret data that should be hidden, where the carrier is the material that takes the message in it [8]. The general model of data hiding can be described as follows. The embedded data is the message that one wishes to send it securely. It is usually hidden in an innocuous message referred to as a cover-text or cover-image or cover-audio as appropriate, producing the stego-text or other stego-object. A stego-key is used to control the hiding process so as to restrict detection and/or recovery of the embedded data to parties who know it. Steganography Techniques Used for Information Hiding for Integrity our information from penetration or alteration [8], [9].

## 2.6 Overview on Compression Techniques of Digital Image Compression technique

In this paper, the JPEG image compression technique was used. JPEG is one of the well-known image compression techniques. The details of JPEG are explained below [5].

1) Discrete Cosine Transform (DCT): The DCT is the first step in JPEG as shown in Figure 1. The goal of this step is to move the lower frequencies reside in the top-left corner of the spectrum, while the higher frequencies are concentrated in, the lower right, i.e. the DC component, i.e. first coefficient, is located at (0, 0), at the upper-left most value, while all the other 63 coefficients are called AC components. The DC component represents the sum of the 64 pixels in the input 8 × 8 pixel block multiplied by the scaling factor.

$$D(u,v) = \left(\frac{4}{\ldots\ldots}\right) \sum_{X=0}^{M-1} \sum_{Y=0}^{N-1} p(X,Y) \left(\cos\left(\frac{\ldots+\ldots}{\ldots}\right) * \cos\left(\frac{\ldots+\ldots}{\ldots}\right)\right) \quad (1)$$

where u is raw number in transformed data, v is column number in transformed data, x is raw number in original data, y is column number in original data, M is number of rows in original data, N is number of columns in original data, D is value of pixel in transformed data, and P is value of pixel in original data.

2) Quantization: In this step, the coefficients of DCT are divided by the corresponding quantize step-size parameter Q(u, v) in the quantization matrix and rounded to the nearest integer as follows, Dq(u, v) = Round((D(u, v))/(Q(u, v))). This step removes the high frequencies present in the original image by dividing values at high indexes in the vector (the amplitudes of higher frequencies) with larger values than the values by which are divided the amplitudes of lower frequencies. As a result of division and rounding process, we will have a lot of consecutive zeroes due to most images have a small quantity of high detail at high frequency.

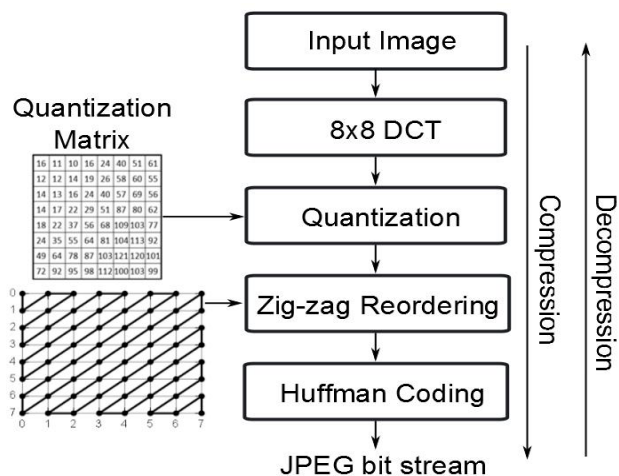

FIG. 1 BLOCK DIAGRAM OF THE JPEG COMPRESSION TECHNIQUE.

3) Zig-zag Reordering: In this step, all AC coefficients are scanned in a Zig-Zag manner (see Figure 1) to traversing the DCT coefficients in the order of increasing the spatial frequencies.

4) Huffman Coding: Huffman coding is used to reduce the content of files or image by a variable-length encoding technique which encodes characters that occur less frequently by more bits while encoding characters that occur more frequently by fewer bits. There are two types of coding methods, namely, fixed length and variable length. However, Huffman encoded file needs to provide a header with the information about the table used, and this header will be used to decode the file. To generate the Huffman tree, first, each character gets a weight equal to the number of times it occurs in the file. Second, we collect smallest weight characters to build large tree and so on. This tree is used in encoding and decoding processes.

## 2.7 Overview on Combined Crypto-Steganography-Compressed approaches

The Combining encryption methods of cryptography and steganography enable the user to transmit information which is masked inside of a file in plain view. This will provide more security to transferring data. Steganography and cryptography are used together to ensure the security of the covert message. In this section, we use combine technique consist of symmetric stream substitution encryption by using Hill cipher algorithm for concealing covered image which contains a hidden text then decrypt cover image to get the original text so that it is computationally infeasible to be interpreted by any eavesdropper.

$$C = PK \mod 26 \quad (2)$$
$$\text{For encryption: } C = E_k(p) = KP[4] \quad (3)$$
$$\text{For decryption: } P = D_k(C) = K^{-1}C = K^{-1}KP = P[4] \quad (4)$$

Where C and P, are row vectors of length 3 representing the plaintext and ciphertext, and K is a matrix representing the encryption key. Operations are performed mod 26. The key transform and hide in the cover image which also has the cipher text hidden in it; so no need secure channel to transmission keys and distribution center. In this research, we use decryption algorithm (Hill cipher algorithm) for plain decrypt data (cover image) which contain hide text (key) for getting cipher data. As the first step, we add the cipher text which is obtained by using Hill cipher technique to the cover image. In the second step, we add the encrypted key into the cover image which forms the encrypted image. The encrypted image is communicated over an unsecured channel. At the receiver side, the key is extracted from the cover image and by using its inverse the plaintext is obtained.

## 3 PROPOSED ALGORITHM FOR COMBINED CRYPTO-INTEGRITY-STEGANOGRAPHY-COMPRESSED APPROACHES

The proposed model is explained in this section. As shown in Figure (2), the proposed model consists of two phases, namely, send and receive.

A. Send phase: In the send phase, the original message is encoded using Cesar method (E). At the same step, the SHA-256 algorithm is used to generate a fixed-length 256-bit hash. The encrypted message (E (m)), private key (P), and the output of hash function (SHA (M)) are concatenated (row by row), to for a three-rows block of data D. The length of the generated message from SHA function is 256; hence, the zero-padding method is applied to the encrypted message and private key to make them with the same length of (SHA (M)). Thus, the size of D is 3×128, i.e. 384 elements. The D is then reshaped to be 8×48, i.e. 48 blocks, and then the DCT is then applied to each block. Next, Zig-zag ordering is used to sort the output of DCT and then Huffman coding is used to generate the compressed text. Finally, in this phase, the compressed message is hidden in the original image.

B. Receive phase: In this phase, the reverse of the steps of send phase is applied. As shown in Figure 2, at the receiver side, the image is received, and the hidden message is retrieved. Next, Huffman decoding, zig-zag ordering, and inverse DCT are applied. Then, the result from IDCT (i.e. inverse DCT) is reshaped to be in the original for, i.e. 3 × 256, and then the encrypted message and the private key are used to decode the message. Finally, the SHA-256 hash function is applied to the decrypted message and then compared with SHA(M) to verify that is the message is changed or not. It is worth mentioning that SHA-256 function is used in our model as a digital signature or password validation.

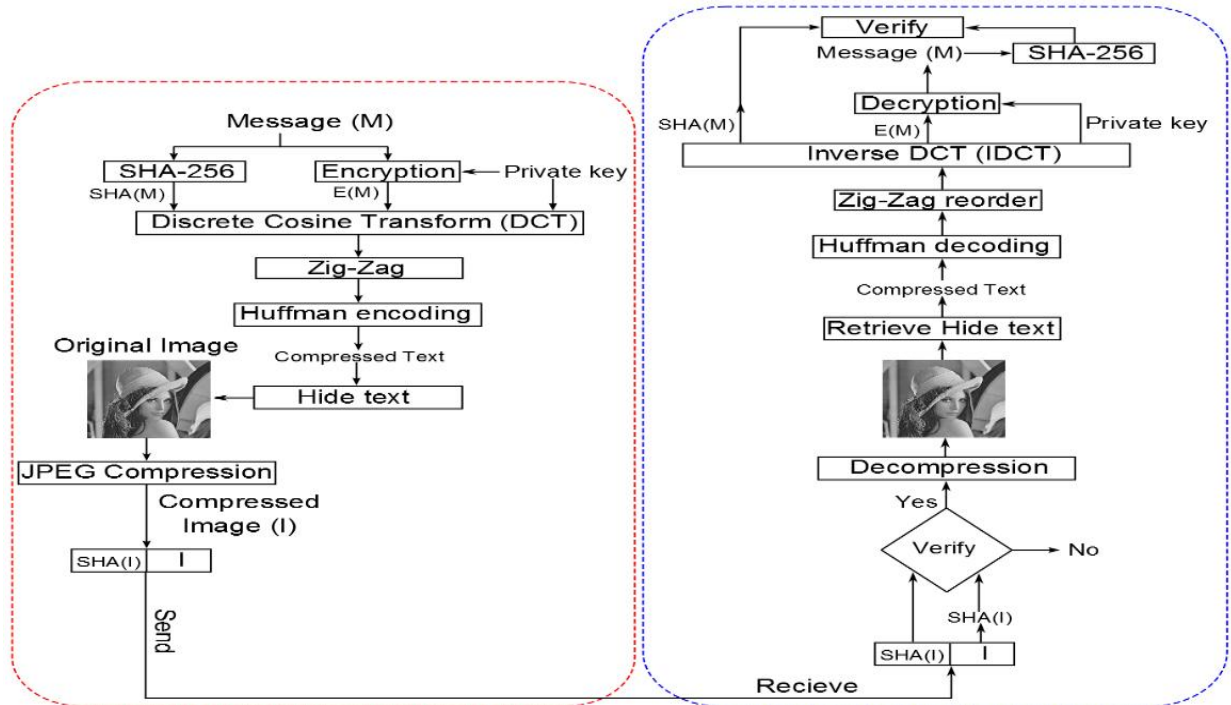

FIG. 2 BLOCK DIAGRAM OF THE PROPOSED MODEL

## 4 SIMULATION RESULTS

In this section, different experiments were conducted to evaluate the proposed model. In the first experiment, the proposed model was evaluated. In this experiment, we try to explain the steps of the proposed model and the inputs and outputs of each step. In the second phase, the influence of the scale of the original image was tested on the proposed model, i.e. on the size of the original message. In the first experiment, the original message was "I'm so proud to be Egyptian", the private key was 16, and the encoded message was "YUUU Ydjuhdqjyedqb Sedvuhudsu ed Secfkjuh Udwyduuhydw qdt coijuci 8YSSUc9". The hash code for the original message was (444429EA64FFA261EB1054E803733B132BBB5136A5944E4D492067A2DCD58F68E127E785D351F569C37CB486F7E0F0029FCA3F8D3F23F47662DE058A73100A63). As mentioned in the proposed model, the encrypted message, the private key, and the output of hash function are concatenated and then reshaped to apply DCT. Next, Zig-zag ordering and Huffman coding are applied, and the length of the compressed message was 188, i.e. Huffman coding compressed 384 elements to 188 elements; hence, the compression ratio was 384 188 = 2.0426. The compressed text is then hidden in the original image in Figure 3. The compressed message needs 1344 pixels, while the original image has 256 × 256 = 65536 pixels. As shown, the hidden message in the upper left corner of Figure 3b can be recognized. Following the steps at the receive phase, the original text can be retrieved and then the hash function is applied to verify that the message was changed or not. Our experiment was conducted on different scales, and the proposed model achieved perfect results. To sum up, the proposed model achieved high integrity protection levels and combined different techniques together, which improved the performance of the proposed model. Moreover, the novelty of this model is another advantage against other standard techniques.

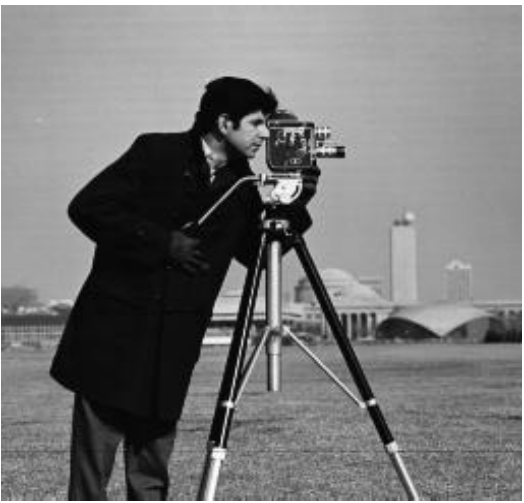
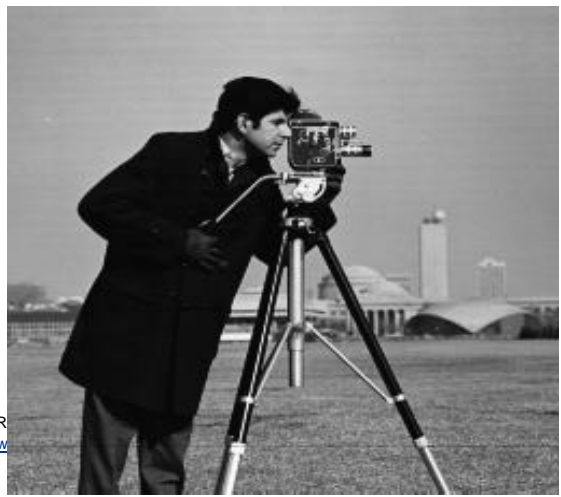

(A)            (B)

FIG. 3. THE ORIGINAL IMAGE AND THE IMAGE THAT CONTAINS THE HIDDEN TEXT IN THE UPPER LEFT CORNER: (A) ORIGINAL IMAGE, (B) THE IMAGE WITH THE HIDDEN TEXT.

## 5 CONCLUSION

The steganography based approach to hide the evidence of originality of the photos on creation is vital nowadays digital Era that is characterized by difficulties to distinguish photos based on integrity or credibility. The problem severity appears more in photos used in official and legal conditions. One solution is to ensure no physical access to the camera, but the challenge is still there if the camera is an IP camera. The simulation results proved that the image could be integrity protected with minimal unobserved impact on the quality of image if the image processing stage posts the capturing stage include the huddling of integrity code including the ID of the camera in the image that becomes the image that we can consider it digitally the original copy.